%% file: main.tex
\documentclass[11pt,a4paper]{article}
\usepackage[T1]{fontenc}
\usepackage[utf8]{inputenc}
\usepackage{newtxtext,newtxmath}
\usepackage[margin=25mm]{geometry}
\usepackage{microtype}
\usepackage{graphicx,booktabs,tabularx,array}
\usepackage{amsmath}
\usepackage{float,placeins}
\usepackage[font=small,labelfont=bf,labelsep=period]{caption}
\usepackage[super,sort&compress,numbers]{natbib}
\usepackage{xcolor}
\definecolor{linkblue}{HTML}{28567A}
\usepackage[colorlinks=true,allcolors=linkblue]{hyperref}
\hypersetup{
 pdftitle={Entropy-Guided Reverse-Causal AI to Identify Upstream Bottleneck Genes for Alzheimer’s Drug Discovery},
 pdfauthor={Victor O. K. Li; Jacqueline C. K. Lam; Yang Han; Lawrence Y. L. Cheung},
 pdfsubject={Proposed methodology for bottleneck-directed drug discovery with a computational feasibility study},
 pdfkeywords={Alzheimer's disease, reverse-causal tracing, network medicine, gene prioritization, entropy}
}
\usepackage{fancyhdr}
\graphicspath{{figures/}}
\newcommand{\gene}[1]{\textup{#1}}
\newcommand{\norm}[1]{\mathcal{N}(#1)}

\title{\vspace{-1.3cm}\bfseries Entropy-Guided Reverse-Causal AI to Identify Upstream Bottleneck Genes for Alzheimer’s Drug Discovery}
\author{\normalsize Victor O. K. Li$^{1,2}$, Jacqueline C. K. Lam$^{1,2,3,4}$,\\[3pt]
\normalsize Yang Han$^2$ and Lawrence Y. L. Cheung$^{2,4}$\\[8pt]
\small $^1$The University of Hong Kong, Hong Kong SAR, China\\
\small $^2$ElohaX, Hong Kong SAR, China\\
\small $^3$University of Cambridge, United Kingdom\\
\small $^4$CASBS, Stanford University, United States}
\date{}

\begin{document}
\maketitle
\thispagestyle{plain}
\begin{abstract}
Identifying upstream regulators that connect several disease processes to therapeutic interventions is a central objective in Alzheimer's disease drug discovery. We propose an entropy-guided reverse-causal framework that makes candidate bottleneck genes the organizing link between disease mechanisms, pathways, molecular targets and drugs. The methodology integrates five stages: an Alzheimer's-specific knowledge graph with language-model assistance and expert review; reverse tracing from drugs to candidate genes; entropy-guided prioritization; forward propagation to drugs and complementary combinations; and staged validation with evidence feedback. The novelty lies in integrating upstream bottleneck identification, entropy-guided prioritization and iterative therapeutic selection within a dynamic, bidirectional discovery architecture. We demonstrate its molecular tracing and gene-prioritization components in a computational feasibility study using DeepDrug2 and MSigDB pathway annotations. Tracing amlodipine, indapamide and atorvastatin through a network of 11,300 molecular and drug nodes identifies 46 routes to nine genes. \gene{EGFR} is the leading candidate, supported by 26 routes from all three drugs; \gene{MME} and \gene{MAF} rank next. These results show how pharmacological starting points can identify shared candidate genes with defined molecular connections. The framework's scientific significance lies in connecting convergent disease mechanisms to systematic intervention selection, with preservation of cognition and independence as the translational objective.
\end{abstract}

\section{Introduction}

Preserving cognition and independence in Alzheimer's disease (AD) motivates the discovery of interventions directed at the mechanisms that initiate and sustain neurodegeneration. AD involves interacting processes in amyloid processing, tau biology, immune responses, lipid metabolism and intracellular trafficking.\cite{knopman2021,bellenguez2022} Upstream regulators and genes at points of pathway convergence offer hypotheses for influencing several of these processes through shared intervention points. Connecting these genes to candidate drugs is a central task for mechanism-oriented therapeutic discovery.

This task brings together two methodological needs: identifying candidate upstream bottlenecks and maintaining an integrated route from those candidates to therapeutic selection. Network medicine represents disease through interacting molecular systems,\cite{barabasi2011} and drug-target proximity and biomedical knowledge graphs supply useful foundations for connecting interventions to disease modules.\cite{guney2016,chandak2023} A discovery methodology can build on these foundations by tracing candidate drugs toward their shared molecular determinants, prioritizing the resulting genes and returning those priorities to the selection of drugs and combinations. As biological and translational evidence accumulates, the same representation can support successive revisions to this process.

We propose an entropy-guided reverse-causal framework for prioritizing upstream bottleneck genes and drugs in AD. Here, \emph{causal} denotes a disease-driving role in the initiation or progression of AD. Its five stages integrate an AD-specific knowledge graph, reverse tracing, entropy-guided prioritization, forward therapeutic selection and staged validation with evidence feedback. A \emph{candidate bottleneck gene} is a potential disease-driving regulator connecting molecular processes to therapeutic intervention points. Reverse-causal tracing follows drug--target and pathway relationships toward these genes. Forward selection uses gene priorities to identify pathways, targets, drugs and complementary combinations, forming an iterative gene--pathway--drug discovery cycle.

The novelty of the proposed methodology lies in three connected design elements. First, reverse-causal reasoning makes upstream bottlenecks the link between candidate interventions and converging disease mechanisms. Second, entropy-guided prioritization combines bottleneck value with the distribution of supporting evidence, making the rationale for gene selection explicit. Third, a dynamic, bidirectional graph couples gene discovery to drug and combination selection and to updates from validation evidence. Language-model assistance and expert review provide a structured means of proposing and checking AD-specific relationships within this graph. The contribution is the integration of these elements into a coherent discovery cycle.

The scientific significance lies in connecting upstream bottleneck discovery to interventions spanning multiple AD-related mechanisms. Upstream regulatory hypotheses guide the search for intervention points early in disease processes. Pathway convergence organizes the selection of drugs and combinations around mechanisms that span several aspects of AD biology. Evidence feedback connects these molecular hypotheses to population, preclinical and clinical investigation. This framework therefore provides a systematic approach to studying interventions intended to preserve cognitive function and support healthy longevity.

We illustrate the central reverse-tracing and gene-prioritization mechanism through a computational feasibility study using the DeepDrug2 biomedical network and human Molecular Signatures Database (MSigDB) annotations. Starting from amlodipine, indapamide and atorvastatin, the study identifies 46 paths to nine AD-focused genes. \gene{EGFR} is the only gene reached from all three drugs and has the highest priority, followed by \gene{MME} and \gene{MAF}. These results demonstrate how the proposed methodology can be applied to an existing molecular network, producing gene hypotheses with explicit supporting paths for subsequent drug selection and validation.

\section{Related Work}

\subsection{Network medicine and mechanism-oriented drug repurposing}

Network medicine represents disease through interacting molecular systems and disease-associated modules.\cite{barabasi2011} Menche and colleagues showed that the relative positions of disease modules in the human interactome capture shared molecular and phenotypic features across diseases.\cite{menche2015} Guney and colleagues subsequently used the proximity between drug targets and disease proteins to support drug efficacy screening.\cite{guney2016} These studies provide a rationale for examining short molecular routes from drugs to disease-related genes and for interpreting shared network neighborhoods as sources of mechanistic hypotheses.

The study of bottlenecks adds a complementary focus on nodes that connect molecular routes. Yu and colleagues related high betweenness to gene essentiality and expression dynamics in yeast protein interaction and regulatory networks.\cite{yu2007} Network control theory further formalizes how selected nodes influence system behavior under a specified network and dynamical model.\cite{liu2011} Our framework operationalizes candidate bottlenecks through drug-linked path convergence, molecular proximity, pathway support and reachability. This formulation directs gene prioritization toward the molecular context of the selected interventions.

Drug repurposing studies also illustrate the value of connecting computational predictions to several forms of biological evidence. Cheng and colleagues combined interactome proximity with longitudinal healthcare data and cellular assays to investigate predicted drug--disease relationships.\cite{cheng2018} Taubes and colleagues integrated transcriptomic repurposing with experimental and real-world evidence in a study of bumetanide and APOE4-related AD.\cite{taubes2021} Such work motivates a discovery architecture in which a computational hypothesis carries a molecular rationale into successive biological and population-level evaluations. Reverse tracing contributes by identifying the candidate genes and connecting routes through which the starting drugs can be investigated.

\subsection{Biomedical knowledge graphs and graph-based drug discovery}

Biomedical knowledge graphs integrate molecular and clinical concepts in a form that supports both representation learning and relationship-level inspection. Hetionet combined heterogeneous biomedical relationships and used network path patterns to prioritize drug repurposing candidates, with individual paths providing pharmacological explanations.\cite{himmelstein2017} PrimeKG brings together multimodal biomedical information for precision medicine,\cite{chandak2023} and TxGNN uses graph learning to support drug repurposing across diseases, including settings with sparse treatment information.\cite{huang2024} Together, these approaches establish a foundation for linking drug predictions to the molecular relationships that support them.

DeepDrug combines expert knowledge, a weighted heterogeneous biomedical graph and graph neural networks to prioritize drugs and drug combinations for AD.\cite{li2025deepdrug} DeepDrug2 extends this research program with a germline-focused graph and drug analyses using electronic health records.\cite{li2025deepdrug2} The proposed framework builds on this trajectory by making candidate bottleneck genes the organizing link between drug discovery, mechanistic interpretation and evidence feedback. The DeepDrug2 graph provides the molecular substrate for demonstrating the reverse-tracing component in the computational feasibility study.

\subsection{Reverse reasoning and pathway-informed gene prioritization}

Reverse reasoning uses observed outcomes or selected interventions to identify candidate upstream molecular determinants. CARNIVAL combines transcription factor and pathway activity estimates from gene expression with signed, directed signaling relationships to infer upstream regulatory hypotheses.\cite{liu2019carnival} Reverse-Gene-Finder traces influential neural representations back to gene tokens to identify candidate genetic biomarkers of AD.\cite{li2025rgf} These approaches illustrate how prior biological knowledge and backward reasoning can connect downstream information to molecular hypotheses. In the proposed methodology, selected drugs provide the starting points, and bounded paths through explicit biomedical relationships identify candidate genes shared across their molecular neighborhoods.

Curated pathway resources provide biological context for these explanations. MSigDB distributes annotated gene sets,\cite{liberzon2011} including KEGG-derived collections that organize molecular functions and disease-related processes.\cite{kanehisa2025} In the proposed architecture, curated knowledge and expert review ground the relationships used for reverse reasoning and subsequent drug selection. The feasibility study uses pathway memberships to describe the biological context of traced molecular entities. The framework brings these ideas together through an explicit cycle from drugs to candidate bottlenecks, from bottlenecks to therapeutic hypotheses and from validation evidence to updated discovery priorities.

\section{Data and Methodology}
\label{sec:methods}

\begin{figure}[H]
\centering
\includegraphics[width=\textwidth]{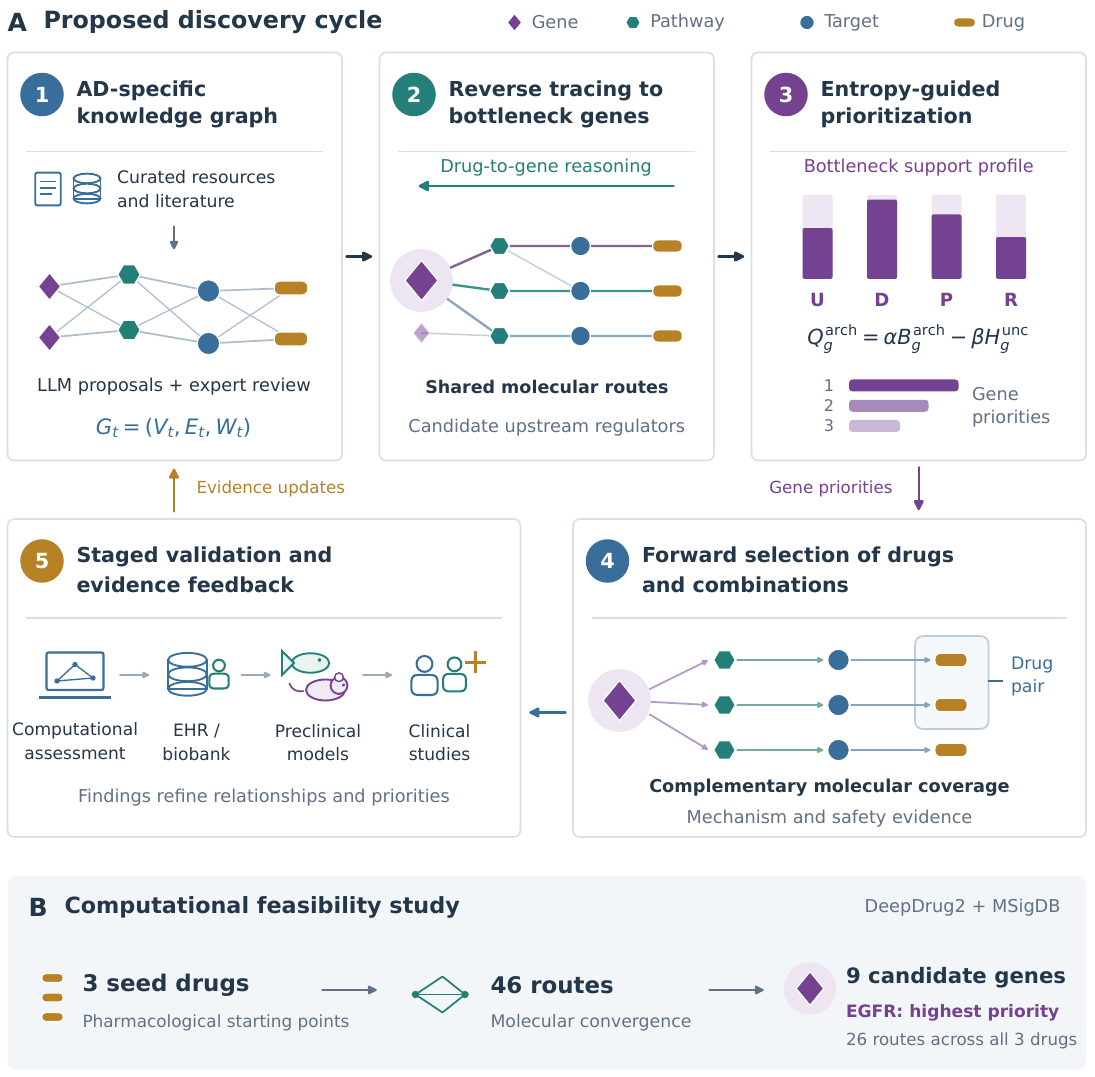}
\caption{\textbf{Proposed five-stage methodology for upstream bottleneck discovery.} (A) Curated knowledge, language-model assistance and expert review build the AD-specific graph. Reverse tracing identifies genes at the convergence of drug-linked molecular routes. Bottleneck support and an entropy-derived term guide gene priorities, which inform forward selection of drugs and complementary combinations. Staged validation feeds evidence back into the graph. Entity colors and shapes are consistent across stages; arrows show the direction of reasoning and evidence flow. Network layouts, support bars and rankings in panel A are schematic. $U$, upstreamness; $D$, disease-driving support; $P$, pathway control; $R$, drug-target reachability. (B) The computational feasibility study integrates DeepDrug2 with MSigDB and traces three drugs through 46 routes to nine genes. \gene{EGFR} has the highest priority and 26 routes spanning all three drugs. LLM, large language model; EHR, electronic health record.}
\label{fig:overview}
\end{figure}

\input{sections/framework}

\subsubsection{Weighted molecular reverse tracing}

Reverse tracing follows drug-to-target relationships, protein-protein interactions in both directions, and protein-to-gene relationships. Each path contains distinct nodes and terminates at the first gene reached. Path weights favor strong interactions and shorter molecular routes. Pathway annotations provide biological context for the traced genes, proteins and targets. Table~\ref{tab:settings} specifies the path-length and search limits used in the feasibility study.

For an edge $e$ of relationship type $r(e)$, its normalized weight is
\begin{equation}
\widetilde{w}_e=\operatorname{clip}_{[0,1]}
\left(\frac{w_e}{\max_{e':r(e')=r(e)}w_{e'}}\right).
\label{eq:edge}
\end{equation}
For a path $\pi$ containing $\ell(\pi)$ edges and starting at drug $d$, its weight is
\begin{equation}
w(\pi)=a_d\,\delta^{\ell(\pi)-1}\prod_{e\in\pi}\widetilde{w}_e,
\qquad a_d=1,\quad\delta=0.85.
\label{eq:path}
\end{equation}
The multiplicative weight summarizes the strength of the participating relationships, while the decay favors shorter routes. Drug-drug associations remain part of the source network; the traversal uses the three molecular relationship types specified above.

\subsubsection{Four-component bottleneck profile}

The bottleneck profile separates four aspects of graph support so that each gene's priority can be interpreted. Let $\Pi_g$ be the set of enumerated paths ending at gene $g$, $W_g=\sum_{\pi\in\Pi_g}w(\pi)$ their total weight, $n_g$ the number of distinct seed drugs reaching $g$, and $q_g$ the number of relationship types represented across these paths. Let $d_g=\min_{\pi\in\Pi_g}\ell(\pi)$ denote the shortest reverse distance. We normalize each raw feature across the reachable gene set $\mathcal{G}$ using
\begin{equation}
\norm{x_g}=\frac{x_g-\min_{h\in\mathcal{G}}x_h}
{\max_{h\in\mathcal{G}}x_h-\min_{h\in\mathcal{G}}x_h}.
\label{eq:norm}
\end{equation}
For a constant raw feature, we assign a normalized value of one when the common value is positive and zero otherwise. All four final components lie in $[0,1]$.

\paragraph{Proximity as an upstreamness proxy.}
We define $U_g=1-\norm{d_g}$. A high value indicates that the gene is reached by a relatively short route from the selected drugs. This measure describes proximity between drugs and genes along the traced paths. The biological interpretation of upstream regulation is a hypothesis attached to prioritized genes and their pathway context.

\paragraph{Weighted convergence support.}
The convergence component rewards path strength, support from distinct drugs and relationship-type coverage:
\begin{equation}
C_g=\mathcal{N}\!\left(W_g[1+0.25(n_g-1)]
[1+0.15(q_g-1)]\right).
\label{eq:convergence}
\end{equation}
Because every scored gene is reached by at least one path, $n_g\geq1$ and $q_g\geq1$. This component summarizes molecular network evidence supporting each candidate gene.

\paragraph{Pathway support and specificity.}
Let $\mathcal{K}_g$ contain the active pathways with at least one membership link to a gene, protein or target on any path in $\Pi_g$. For pathway $k$, let $m_k$ be its source membership count, $c_{gk}$ the number of distinct traced nodes belonging to it, and $I_{gk}$ indicate direct membership of gene $g$. Its specificity and contribution are
\begin{align}
\sigma_k&=\min\!\left(1,\sqrt{100/m_k}\right),\\
A_{gk}&=\omega_k\sigma_k\log_2(1+c_{gk})
\left(1+0.25I_{gk}\right),
\label{eq:pathways}
\end{align}
where $\omega_k=1$ for the active KEGG MEDICUS collection. Membership counts operate on graph nodes, preserving the representation of genes and proteins as separate entities. Let $A_g=\sum_{k\in\mathcal{K}_g}A_{gk}$ and let $\overline{\sigma}_g$ be the contribution-weighted mean of $\sigma_k$. The pathway component is
\begin{equation}
P_g=\mathcal{N}\!\left(A_g f_g\log_2(2+|\mathcal{K}_g|)\right),
\qquad f_g=1-0.20(1-\overline{\sigma}_g).
\label{eq:pscore}
\end{equation}
For an empty pathway set, $A_g=0$ and $f_g=1$. The component combines functional coverage, traced-node overlap and a specificity adjustment for large sets. The KEGG sets used in this analysis all contain fewer than 100 members, so $\sigma_k=f_g=1$ throughout this experiment.

\paragraph{Drug-target reachability.}
We define $R_g=\mathcal{N}(W_g\log_2(1+n_g))$. This component favors strong connections to several seed drugs. Both $C_g$ and $R_g$ depend on weighted paths; their different transformations emphasize convergence structure and pharmacological access, respectively. They are related summaries of the same molecular network.

The composite bottleneck profile is
\begin{equation}
B_g=0.30U_g+0.25C_g+0.25P_g+0.20R_g.
\label{eq:bottleneck}
\end{equation}
The fixed weights give proximity the largest individual contribution and distribute the remaining weight across molecular convergence, pathway support and reachability. These coefficients specify the prioritization rule used in the feasibility study.

\subsubsection{Entropy-guided priority scoring}
\label{sec:feasibility-entropy}

The feasibility study specifies the entropy-derived uncertainty term in Eq.~\ref{eq:architecture-priority} through the balance among the positive graph-derived components that represent the four proposed bottleneck dimensions. Write $x_g=(U_g,C_g,P_g,R_g)$, let $J_g=\{j:x_{gj}>0\}$ and let $k_g=|J_g|$. For $k_g\geq2$, define
\begin{equation}
p_{gj}=\frac{x_{gj}}{\sum_{i\in J_g}x_{gi}},
\qquad E_g=-\frac{\sum_{j\in J_g}p_{gj}\log p_{gj}}{\log k_g},
\qquad H_g=1-E_g.
\label{eq:entropy}
\end{equation}
For fewer than two positive components, set $H_g=1$. High component entropy $E_g$ corresponds to balanced positive support and produces a small penalty $H_g$. Concentration in one component increases the penalty. The normalization uses the number of positive components; consequently, $H_g$ describes their relative balance, while $B_g$ supplies the magnitude and coverage information across all four dimensions.

Using $H_g$ as the study-specific value of $H_g^{\mathrm{unc}}$ in Eq.~\ref{eq:architecture-priority}, the uncertainty-adjusted score and its reported rescaling are
\begin{equation}
T_g=B_g-0.25H_g,
\qquad S_g=\frac{T_g+0.25}{1.25}.
\label{eq:priority}
\end{equation}
Since $B_g,H_g\in[0,1]$, the algebraic bounds $T_g\in[-0.25,1]$ map to $S_g\in[0,1]$. The affine transformation preserves the ranking. $S_g$ is a relative network-support index under the specified graph, seed set and normalization. $H_g$ is a component-balance heuristic; the two quantities summarize the strength and balance of graph support, respectively.

\subsubsection{Gene rankings and feasibility assessment}

We assess computational feasibility through the genes reached from the selected drugs, the number and strength of their supporting molecular routes, and convergence across distinct drugs. Gene priorities are calculated from all enumerated paths using Eqs.~\ref{eq:bottleneck}--\ref{eq:priority}. Comparing the four components and the entropy-derived penalty explains which forms of molecular and pathway support determine each gene's rank. The analysis describes one molecular network and one prespecified seed set.

\FloatBarrier
\section{Experimental Results}

The computational feasibility study examines whether the three selected drugs converge on shared AD-related genes and how molecular and pathway evidence determine their priorities. The resulting gene hypotheses provide a basis for subsequent drug selection and biological investigation.

\subsection{Study design and data sources}

The study combines the DeepDrug2 molecular network, human pathway annotations and three prespecified drug seeds. AD-focused genes in DeepDrug2 define the candidate gene set, while pathway memberships provide the biological context for evaluating convergence among drug-linked molecular routes.

The DeepDrug2 network contains 11,300 nodes: 454 drugs, 1,477 molecular targets, 9,298 other proteins and 71 AD-focused genes. Its 275,348 relationships comprise 7,448 drug-target interactions, 195,883 protein-protein interactions, 71 gene-protein links and 71,946 drug-drug associations. The analysis uses the relationship types, weights and directions provided by DeepDrug2. Genes and their encoded proteins remain distinct node types, linked through gene-protein relationships. The 71 gene nodes define the candidate gene set for this experiment.

We augment this network with human MSigDB v2026.1.Hs pathway memberships.\cite{liberzon2011,msigdb2026} Symbol-based memberships are matched to gene, protein and target labels; existing gene-protein links also support the mapping. Pathways are generally retained when at least two members map to the graph and the source set contains at most 500 members. The integrated graph adds 3,379 pathway nodes and 115,012 membership relationships. These comprise 648 KEGG MEDICUS, 1,807 Reactome and 924 WikiPathways sets. The resulting graph therefore contains 14,679 nodes and 390,360 relationships.

KEGG MEDICUS provides the pathway annotations used for gene scoring and biological interpretation. Reactome and WikiPathways memberships are included in the integrated graph as additional sources of pathway knowledge. Molecular interactions determine the drug-to-gene routes; pathway memberships characterize the biological processes represented along those routes and contribute to gene prioritization.

\subsection{Seed selection and analysis parameters}

Amlodipine (DB00381), indapamide (DB00808) and atorvastatin (DB01076) were selected from the DeepDrug2 discovery context before tracing. Each drug receives an initial weight of one. Equal weighting allows the analysis to compare the molecular support associated with each drug under the same scoring rule.

For each drug, we examine molecular paths containing at most four interactions. At each node, the 40 strongest eligible connections are considered, ranked by normalized interaction weight. The search retains up to 2,000 partial paths with the greatest cumulative weight at each depth and 5,000 gene-reaching paths per drug. These limits define the molecular neighborhood examined around each seed. A path ends when it reaches a candidate gene.

Gene priorities combine molecular routes across the three drugs with KEGG pathway support. The component weights are 0.30, 0.25, 0.25 and 0.20 for proximity, convergence, pathway support and reachability, respectively; the entropy penalty weight is 0.25. All enumerated routes contribute to scoring. Table~\ref{tab:settings} summarizes the analysis parameters.

\begin{table}[!tb]
\centering\small
\caption{\textbf{Analysis parameters for the feasibility study.} All enumerated gene-reaching paths contribute to gene priorities.}
\label{tab:settings}
\begin{tabularx}{\textwidth}{@{}lX@{}}
\toprule
Parameter & Setting \\
\midrule
Seed drugs & Amlodipine, indapamide, atorvastatin; equal weight \\
Active pathway collection & KEGG MEDICUS \\
Maximum path length & 4 interactions \\
Neighbors retained per node & 40, ordered by normalized edge weight \\
Partial / completed-path limits & 2,000 partial paths per depth / 5,000 paths per seed \\
Protein interaction traversal & Both directions \\
Path-length decay & $\delta=0.85$ after the first interaction \\
Component weights & $U:0.30$, $C:0.25$, $P:0.25$, $R:0.20$ \\
Entropy-derived penalty weight & $\beta=0.25$ \\
Pathway specificity & Reference size 100; penalty weight 0.20 \\
\bottomrule
\end{tabularx}
\end{table}

\subsection{Reverse tracing connects three drugs to nine genes}

\begin{figure}[!tb]
\centering
\includegraphics[width=.98\textwidth]{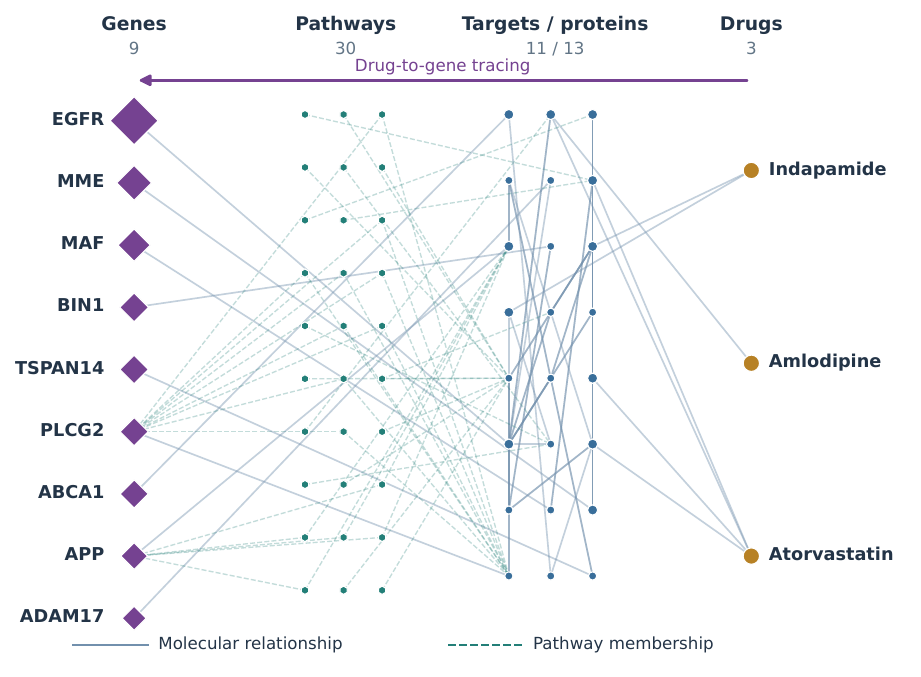}
\caption{\textbf{Molecular routes connecting the seed drugs to candidate genes.} Drugs on the right connect through molecular targets and proteins to genes on the left. Solid links represent molecular relationships; dashed links indicate pathway memberships. Gene marker size increases with priority. The illustrated network contains 66 nodes and 106 relationships, including 30 selected KEGG pathways. For legibility, up to 25 paths per gene are shown, yielding 45 of the 46 routes used for scoring; 25 of the 26 routes to \gene{EGFR} are included. Pathways provide biological context for the traced molecular entities.}
\label{fig:network}
\end{figure}

Reverse tracing identified 46 molecular routes from the three seed drugs to nine genes (Table~\ref{tab:ranking}; Fig.~\ref{fig:network}). The candidates differ in their supporting drugs, path lengths and pathway associations. This variation provides a basis for distinguishing shared molecular convergence from gene connections associated with individual drugs.

\begin{table}[!tb]
\centering\small
\caption{\textbf{Candidate bottleneck genes ranked by network-support priority.} Paths are all enumerated routes to a gene; drugs are distinct starting drugs. Distance is the shortest route in interactions. Pathway counts refer to all connected active KEGG annotations used for scoring. The seed labels identify the drugs supporting each candidate.}
\label{tab:ranking}
\begin{tabular}{@{}rlrrrrrl@{}}
\toprule
Rank & Gene & $S_g$ & Paths & Drugs & Distance & Pathways & Seeds$^\dagger$\\
\midrule
1 & \gene{EGFR} & 1.0000 & 26 & 3 & 3 & 52 & Aml, Ind, Ato\\
2 & \gene{MME} & 0.4210 & 6 & 1 & 3 & 1 & Ato\\
3 & \gene{MAF} & 0.3267 & 2 & 1 & 3 & 2 & Ato\\
4 & \gene{BIN1} & 0.2191 & 4 & 1 & 4 & 1 & Ato\\
5 & \gene{TSPAN14} & 0.2048 & 2 & 1 & 4 & 1 & Ato\\
6 & \gene{PLCG2} & 0.1986 & 2 & 1 & 4 & 47 & Ind\\
7 & \gene{ABCA1} & 0.1971 & 1 & 1 & 4 & 1 & Ato\\
8 & \gene{APP} & 0.1565 & 2 & 1 & 4 & 13 & Ato\\
9 & \gene{ADAM17} & 0.0825 & 1 & 1 & 4 & 39 & Ind\\
\bottomrule
\end{tabular}
\par\smallskip\parbox{.98\textwidth}{\footnotesize $^\dagger$Aml: amlodipine; Ind: indapamide; Ato: atorvastatin. Priority is calculated with Eq.~\ref{eq:priority}. Pathway counts can overlap between genes.}
\end{table}

\FloatBarrier
The recovered genes span distinct patterns of pharmacological access. Atorvastatin reaches \gene{EGFR}, \gene{MME}, \gene{MAF}, \gene{BIN1}, \gene{TSPAN14}, \gene{ABCA1} and \gene{APP}. Indapamide reaches \gene{EGFR}, \gene{PLCG2} and \gene{ADAM17}, while amlodipine reaches \gene{EGFR}. This structure places \gene{EGFR} at the intersection of all three seed-specific neighborhoods and identifies additional candidates associated with individual drugs.

\subsection{EGFR provides the strongest cross-drug convergence}

\gene{EGFR} ranks first with $S_g=1.0000$ and 26 supporting paths, accounting for 56.5\% of the 46 enumerated routes. It is the only candidate reached from all three drugs. Its shortest reverse distance is three interactions, and its traced neighborhood connects to 52 KEGG pathway annotations. These include EGF--JAK--STAT and EGFR--PI3K signaling contexts. All 52 annotations contribute to the pathway-support component for \gene{EGFR}.

The strongest molecular routes identify distinct connections to \gene{EGFR}. Amlodipine and atorvastatin connect through ABCB1 and the EGFR target, whereas indapamide connects through CA9 and the EGFR target. Each route reaches the \gene{EGFR} gene in three interactions, illustrating convergence from different pharmacological starting points.

\begin{figure}[!tb]
\centering
\includegraphics[width=\textwidth]{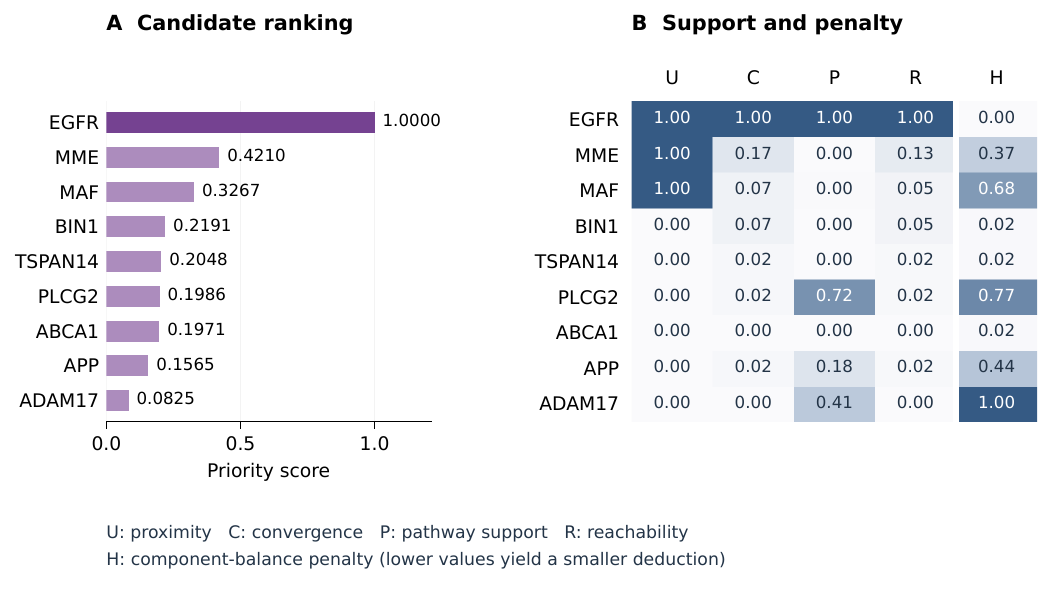}
\caption{\textbf{Gene priorities and their component profiles.} (A) Priorities calculated in the feasibility study using Eq.~\ref{eq:priority}. (B) Normalized proximity ($U$), convergence ($C$), pathway support ($P$), reachability ($R$) and the entropy-derived penalty ($H$). Darker cells indicate larger values. Higher $U,C,P,R$ increase the bottleneck profile, while higher $H$ increases the deduction. Heatmap values are rounded to two decimal places; the calculation uses full precision. A zero after min--max normalization identifies the minimum raw feature value among reachable genes and can coexist with observed molecular or pathway support.}
\label{fig:profiles}
\end{figure}

\gene{EGFR} achieves $U_g=C_g=P_g=R_g=1$ under normalization across the nine reachable genes. It therefore has $B_g=1$, equal positive component proportions and $H_g=0$ (Fig.~\ref{fig:profiles}). The resulting score expresses its leading position on each of the four network features in this experiment. Its priority follows from the joint presence of short paths, strong convergence, pathway coverage and access from multiple starting drugs.

\subsection{Secondary candidates reveal distinct support profiles}

\gene{MME} and \gene{MAF} rank second and third with priorities of 0.4210 and 0.3267. Both are reached from atorvastatin with a shortest distance of three interactions and consequently have $U_g=1$. \gene{MME} has six routes and a weighted path sum of 1.3838; \gene{MAF} has two routes and a weighted path sum of 0.6593. Their convergence components are 0.1706 and 0.0693, respectively. These values account for the stronger molecular support of \gene{MME} for the selected seed drugs.

\gene{BIN1} has four routes and a weighted path sum of 0.6615, close to that of \gene{MAF}. Its shortest distance is four interactions, giving $U_g=0$ after the experiment's distance normalization. The final priorities of 0.2191 for \gene{BIN1} and 0.3267 for \gene{MAF} illustrate how proximity changes the ranking of genes with similar total path support. In this analysis, shortest distances take only the values three and four, so the normalized proximity component separates the genes into two groups.

\subsection{Component balance explains the contribution of pathway support}

\gene{PLCG2} combines 47 connected KEGG annotations with two routes from indapamide. Its pathway component is 0.7192, whereas its convergence and reachability components are 0.0248 and 0.0185. This concentration yields $H_g=0.7660$. The profile score of 0.1897 becomes an adjusted score of $-0.0018$ and a reported priority of 0.1986. This decomposition makes the effect of the entropy-guided rule explicit: substantial pathway coverage contributes positively, and imbalance among the positive components incurs a corresponding deduction.

\gene{ADAM17} has 39 connected pathway annotations and a pathway component of 0.4124. Its other normalized components equal zero, producing $H_g=1$ and $S_g=0.0825$. \gene{APP}, with 13 connected annotations and two routes from atorvastatin, receives a priority of 0.1565. Together, these examples show how the scoring procedure converts heterogeneous graph features into distinguishable hypotheses. The resulting order can be inspected at the level of the individual components that generate it.

\FloatBarrier
\section{Discussion}

\subsection{Upstream bottlenecks as the organizing principle of discovery}

The proposed methodology organizes AD therapeutic discovery around candidate upstream bottlenecks and the mechanisms they connect. Its central contribution is a complete reasoning cycle: integrate disease-specific knowledge, trace candidate interventions toward genes, prioritize the resulting hypotheses, propagate those priorities toward drugs and combinations, and update the graph with validation evidence. This architecture connects the identification of potential disease-driving regulators to therapeutic selection within a shared representation. The same gene hypothesis can therefore be followed from its molecular rationale through successive decisions about intervention and evaluation.

The five stages address complementary parts of this task. Language-model assistance and expert review support the construction of sourced AD-specific relationships. Reverse tracing exposes convergence among pharmacological starting points. Entropy-guided prioritization makes the distribution of support visible alongside its magnitude. Forward propagation uses the resulting genes to organize drug and combination hypotheses, while evidence feedback allows those hypotheses to evolve. The novelty lies in this integration around bottleneck discovery and its bidirectional connection to therapeutic selection.

The computational feasibility study identifies molecular convergence among the three seed drugs. Across the nine genes reached by 46 routes, \gene{EGFR} is supported by 26 routes connecting all three drugs. This shared molecular neighborhood motivates investigation of \gene{EGFR} as a candidate bottleneck. Its intermediate proteins and pathway associations define specific relationships for subsequent pharmacological and biological evaluation.

\subsection{Biological interpretation of prioritized genes}

The leading position of \gene{EGFR} supports focused investigation of growth-factor signaling in the drug-linked network. Prior experimental work identified EGFR as a target in models of amyloid-$\beta$-induced memory impairment.\cite{wang2012} The present result adds a network-based observation: the selected drugs converge on \gene{EGFR}, and the traced neighborhood carries EGF--JAK--STAT and EGFR--PI3K annotations. Together, these observations motivate experiments that relate the implicated molecular routes to cell-specific signaling and AD-relevant phenotypes.

The secondary candidates provide complementary biological entry points. \gene{MME} encodes neprilysin, whose role in brain amyloid-$\beta$ metabolism has been established experimentally.\cite{iwata2001} Its six atorvastatin-linked routes and short reverse distance make amyloid clearance a concrete interpretation to investigate. \gene{MAF} offers a second short-route hypothesis supported by a distinct traced neighborhood.

For \gene{PLCG2}, the strong pathway component directs attention to the signaling relationships connecting indapamide to this gene. Human genetic studies associated the P522R variant in \gene{PLCG2} with reduced AD risk and implicated microglial innate immunity in disease biology.\cite{sims2017} In microglia derived from human induced pluripotent stem cells, heterozygous expression of this variant increased amyloid-$\beta$ clearance while preserving synapses.\cite{solomon2022} These findings motivate testing whether the traced molecular routes influence PLCG2-dependent microglial function, with amyloid clearance and synaptic preservation as relevant biological outcomes.

The recovery of \gene{BIN1} and \gene{ABCA1} connects the drug-linked neighborhood to additional aspects of AD biology. Chapuis and colleagues linked increased \gene{BIN1} expression to AD genetic risk and tau pathology.\cite{chapuis2013} Deletion of \gene{Abca1} increased amyloid deposition in PDAPP transgenic mice, supporting investigation of lipid transport in relation to amyloid accumulation.\cite{wahrle2005} These studies supply biological context for evaluating the traced \gene{BIN1} and \gene{ABCA1} relationships and broaden the mechanistic questions arising from the feasibility study.

Pathway annotations guide interpretation at the level of their recorded membership. Several KEGG MEDICUS labels describe specific variants, exposures or disease settings.\cite{kanehisa2025} In this analysis, their contribution indicates that traced entities occur in the annotated molecular systems. Single-nucleus transcriptomic analysis of human AD brain tissue has revealed disease-associated changes that differ across neuronal and glial cell types.\cite{mathys2019} This cellular diversity motivates evaluating prioritized relationships within the cell types in which their molecular partners are expressed, then relating perturbation responses to the implicated pathways and AD-relevant phenotypes.

\subsection{What the score decomposition reveals}

The four-component representation makes the ranking's assumptions visible. \gene{EGFR} leads across all dimensions; \gene{MME} and \gene{MAF} benefit from short distances; and \gene{PLCG2} carries strong pathway-centered support. These patterns help explain why path counts, pathway counts and final priorities differ. They also show why a single score is most informative when accompanied by its underlying profile.

The entropy term derives from Shannon entropy,\cite{shannon1948} applied to the relative contributions of positive normalized components. The resulting penalty decreases as these contributions become more balanced. Balanced values can occur at either high or low absolute magnitude, so the weighted profile $B_g$ remains essential for interpreting $H_g$. In addition, convergence and reachability share weighted-path information. Their agreement reflects consistency between two transformations of the network evidence. This reading keeps the meaning of the priority tied to its mathematical construction and makes the role of each design choice explicit.

The priorities are conditional on the selected drugs, graph coverage and normalization set. The three-versus-four interaction distances create a discrete proximity contribution in this case, while min--max normalization assigns a zero to the lowest observed pathway support even when one pathway is connected. These observations provide concrete targets for subsequent sensitivity analyses: varying the seed set, depth, neighborhood limit, component weights and annotation collection can quantify the stability of individual hypotheses. Network proximity studies have used reference sets matched for size and node degree to evaluate drug--disease relationships.\cite{guney2016,cheng2019} Analogous controls can characterize the enrichment of observed drug-to-gene convergence relative to network connectivity.

\subsection{Bidirectional discovery and translational significance}

The bidirectional design connects a candidate gene to both the interventions that revealed it and the interventions selected to investigate it. Forward propagation can connect prioritized genes back to pathways, targets and candidate drugs, with combination selection organized around complementary molecular coverage. Cheng and colleagues found that effective combinations for hypertension and cancer were associated with a network configuration in which both drugs reached the disease module while targeting separate neighborhoods.\cite{cheng2019} This finding motivates examining complementarity alongside convergence: shared routes identify a common candidate bottleneck, while complementary routes identify additional mechanisms that a combination could cover. Keeping both views available can guide the formulation of pharmacological hypotheses.

Evidence feedback provides a means to refine these hypotheses as new evidence is incorporated into the graph. Perturbation studies can characterize the direction and cell context of prioritized gene effects. EHR and biobank analyses can evaluate drug-outcome relationships within defined populations, while preclinical studies can test molecular and phenotypic responses. The combined computational, experimental and population analyses reported by Cheng and colleagues and by Taubes and colleagues provide precedents for connecting these forms of evidence in drug repurposing.\cite{cheng2018,taubes2021} In the proposed framework, interpreting new findings in relation to the traced molecular routes can refine the evidence supporting each gene priority. The feasibility study demonstrates the computational procedure for reverse tracing within this discovery cycle. Its broader scientific purpose is to organize early intervention hypotheses around upstream regulation and convergent disease processes. Linking those hypotheses to evidence on cognition and daily function provides a translational direction toward preserving independence and supporting healthy longevity.

\section{Conclusion}

We propose an entropy-guided reverse-causal methodology that makes candidate upstream bottleneck genes the organizing link between AD mechanisms and therapeutic discovery. Its five-stage architecture integrates an AD-specific knowledge graph with language-model assistance and expert review, reverse tracing, entropy-guided prioritization, forward drug and combination selection, and staged validation with evidence feedback. The methodological contribution is a dynamic, bidirectional discovery cycle in which candidate interventions reveal gene hypotheses and those hypotheses guide subsequent therapeutic selection.

A computational feasibility study demonstrates the reverse-tracing and prioritization procedure. Starting from amlodipine, indapamide and atorvastatin, the analysis recovers 46 molecular routes to nine genes. \gene{EGFR} is the leading candidate and the sole gene reached from all three drugs, followed by \gene{MME} and \gene{MAF}. Explicit paths, pathway annotations and component scores explain how these priorities arise. These findings establish a concrete computational foundation for the proposed methodology and identify focused molecular hypotheses for its subsequent stages. By connecting upstream regulatory hypotheses, convergent pathways and iterative intervention selection, the framework provides a systematic approach to AD discovery directed toward preserving cognition and independence.

\section*{Data availability}
The biomedical graph data are available in the supplementary data of DeepDrug2.\cite{li2025deepdrug2} Pathway annotations are available from the human MSigDB v2026.1.Hs collections.\cite{msigdb2026}

\section*{Acknowledgements}
The underlying research program was supported by four US National Academy of Medicine Healthy Longevity Catalyst Awards (2021--2024). We gratefully acknowledge Professors David C. Rubinsztein and James Rowe of the University of Cambridge, and Professor W. C. Chan of Queen Mary Hospital, Hong Kong, for their collaboration and support in this research.


\clearpage
\bibliographystyle{unsrtnat}
\small
\bibliography{references}
\end{document}

%% file: sections/framework.tex
\subsection{Methodological objective and discovery cycle}

The proposed methodology seeks genes with potential to regulate several AD-relevant processes and connects those genes to therapeutic interventions. Here, \emph{causal} denotes a disease-driving role in the initiation or progression of AD. \emph{Upstream} describes regulatory position within a disease process, and \emph{bottleneck} describes a point at which several molecular mechanisms converge. Reverse-causal tracing seeks candidate disease-driving genes by reasoning backward from drugs through their molecular relationships. The complete architecture links this reasoning to forward therapeutic selection and iterative evidence updates (Fig.~\ref{fig:overview}).

We represent the graph at iteration $t$ as $G_t=(V_t,E_t,W_t)$, where nodes encode biomedical entities, edges encode typed relationships and $W_t$ contains their weights. A discovery cycle produces candidate gene priorities, linked pathway and target hypotheses, and an ordered set of drugs or combinations for investigation. The following five stages specify the proposed methodology. Section~\ref{sec:feasibility-method} then describes the computational formulation used to evaluate molecular reverse tracing and gene prioritization.

\subsection{Stage 1: AD-specific knowledge integration with expert review}

The graph provides a shared representation of genes, proteins, pathways, molecular targets and drugs. The proposed design also accommodates AD and multimorbidity phenotypes, genetic evidence and validation findings. These entities connect upstream molecular hypotheses to cognitive and functional outcomes through explicit, sourced relationships. Relationship types, directions and weights describe the molecular connections, while the supporting evidence provides a basis for evaluating their biological relevance.

Knowledge integration combines curated resources with language-model assistance and expert review. Curated molecular and pathway resources, including STRING, KEGG, Reactome and Gene Ontology, provide the biological foundation. Language-model-assisted literature analysis proposes candidate gene--pathway--drug relationships for expert evaluation. Verification against the original sources and domain-expert review determine which proposed relationships enter the graph and how their biological context is represented. Together, literature evidence and expert evaluation establish the biological basis of the relationships used for inference.

New molecular and therapeutic evidence can refine the biological relationships represented in the graph, their weights and the interpretation of implicated pathways. Successive analyses therefore reassess gene and drug priorities in light of the evidence accumulated during validation.

\subsection{Stage 2: Reverse-causal tracing to candidate bottleneck genes}

Reverse tracing starts from selected drug candidates and follows their molecular targets and pathway connections toward candidate genes. Its scientific objective is to identify potential upstream regulators that connect several disease mechanisms to these pharmacological starting points. The molecular routes connecting each candidate to the starting drugs identify points of convergence and the biological processes implicated in those connections.

Bottleneck prioritization combines four dimensions from the proposed methodology: upstream position ($U_g^{\mathrm{arch}}$), disease-driving support ($D_g^{\mathrm{arch}}$), pathway control ($P_g^{\mathrm{arch}}$) and drug-target reachability ($R_g^{\mathrm{arch}}$). At the architectural level, the composite profile is
\begin{equation}
B_g^{\mathrm{arch}}=f\!\left(U_g^{\mathrm{arch}},D_g^{\mathrm{arch}},
P_g^{\mathrm{arch}},R_g^{\mathrm{arch}}\right).
\label{eq:architecture-profile}
\end{equation}
Upstream position concerns a gene's regulatory relationship to disease processes. Disease-driving support integrates evidence relevant to that role. Pathway control concerns the processes influenced by the candidate, while reachability connects those processes to druggable intervention points. For computational evaluation, each dimension is expressed through measurable features derived from the available graph evidence. The feasibility study below uses explicit molecular paths and pathway annotations to construct such a profile.

\subsection{Stage 3: Entropy-guided bottleneck prioritization}

Entropy-guided prioritization couples the magnitude of bottleneck support to an account of its uncertainty. The proposed architecture uses the general form
\begin{equation}
Q_g^{\mathrm{arch}}=\alpha B_g^{\mathrm{arch}}-\beta H_g^{\mathrm{unc}},
\qquad \alpha,\beta\geq0,
\label{eq:architecture-priority}
\end{equation}
where $H_g^{\mathrm{unc}}$ is an entropy-derived uncertainty term defined for the evidence representation. This formulation makes the choice of evidence distribution and the interpretation of uncertainty part of the method specification. When a distribution describes alternative mechanistic explanations, Shannon entropy measures dispersion across those alternatives. When a distribution describes contributions from scoring components, its entropy describes the balance of component support.

The role of this stage is to prioritize candidates with substantial bottleneck support and a well-characterized evidence profile. The contribution of each scoring component and its supporting molecular relationships explains the basis of a gene priority. In the computational feasibility study, $H_g^{\mathrm{unc}}$ is defined as one minus normalized entropy over positive graph-support components. Section~\ref{sec:feasibility-entropy} gives the exact definition used to obtain the reported ranking.

\subsection{Stage 4: Forward propagation to drugs and combinations}

Forward propagation connects the updated gene priorities to pathways, molecular targets and drugs. A candidate bottleneck with increased priority raises the relevance of the mechanisms and intervention points linked to it within the discovery process. This produces a bidirectional reasoning cycle: candidate drugs guide gene discovery, and the resulting gene hypotheses guide subsequent therapeutic selection.

Combination selection is organized around coverage of several prioritized bottlenecks and complementary mechanisms. Shared routes identify mechanisms targeted by several drugs, while distinct routes identify additional coverage contributed by individual candidates. Pharmacological action, biological context and safety evidence guide the interpretation of this coverage. The proposed iteration continues until rankings stabilize or a prespecified iteration limit is reached. This stage translates an upstream regulatory hypothesis into a structured set of intervention hypotheses for evaluation.

\subsection{Stage 5: Staged validation and evidence feedback}

Validation is organized as a sequence from computational assessment to population, preclinical and clinical evidence. Computational analyses examine the stability of gene and drug priorities across graph and scoring choices. EHR and biobank studies investigate drug-outcome relationships within defined populations. Zebrafish and mouse models provide settings for testing molecular and phenotypic responses, and clinical studies connect selected hypotheses to human outcomes. The choice of study at each stage follows the biological question attached to a candidate and its supporting routes.

Evidence feedback completes the discovery cycle. Findings are interpreted in relation to the molecular mechanism, experimental context and phenotype under investigation. Supporting findings can strengthen a proposed relationship or refine its weight; discordant findings can revise the mechanistic hypothesis and redirect prioritization. Updating $G_t$ to $G_{t+1}$ initiates another cycle of reverse tracing and forward propagation. Successive evaluations therefore refine the relationship between the proposed molecular mechanism and the intervention under study.

\subsection{Computational formulation for the feasibility study}
\label{sec:feasibility-method}

We evaluate the framework's reverse-tracing and gene-prioritization stages through a deterministic computational case study. The analysis integrates an existing molecular network with pathway annotations, reconstructs drug-to-gene routes and computes entropy-guided priorities. Its experimental objective is to identify shared gene hypotheses supported by molecular routes and pathway associations. Within this study, a candidate bottleneck is identified through convergence of drug-linked paths and pathway support.

In the computational feasibility study, the proposed upstreamness, disease-driving support, pathway control and drug-target reachability dimensions are represented by drug-to-gene proximity ($U_g$), weighted molecular convergence ($C_g$), pathway support and specificity ($P_g$), and drug-target reachability ($R_g$), respectively. These graph-derived components provide computational proxies for the four dimensions in Eq.~\ref{eq:architecture-profile}. In particular, $C_g$ represents the disease-driving-support dimension $D_g^{\mathrm{arch}}$ through convergence of drug-linked molecular routes on a candidate gene. The four components enter both the composite bottleneck profile and the component distribution used for entropy-guided prioritization, as defined below.

%% file: references.bib
@article{knopman2021,
  author = {Knopman, David S. and Amieva, H{\'e}l{\`e}ne and Petersen, Ronald C. and Ch{\'e}telat, Ga{\"e}l and Holtzman, David M. and Hyman, Bradley T. and Nixon, Ralph A. and Jones, David T.},
  title = {{Alzheimer} disease},
  journal = {Nature Reviews Disease Primers}, volume = {7}, pages = {33}, year = {2021},
  doi = {10.1038/s41572-021-00269-y}
}

@article{bellenguez2022,
  author = {Bellenguez, C{\'e}line and K{\"u}{\c{c}}{\"u}kali, Fahri and Jansen, Iris E. and others},
  title = {New insights into the genetic etiology of {Alzheimer's} disease and related dementias},
  journal = {Nature Genetics}, volume = {54}, pages = {412--436}, year = {2022},
  doi = {10.1038/s41588-022-01024-z}
}

@article{barabasi2011,
  author = {Barab{\'a}si, Albert-L{\'a}szl{\'o} and Gulbahce, Natali and Loscalzo, Joseph},
  title = {Network medicine: a network-based approach to human disease},
  journal = {Nature Reviews Genetics}, volume = {12}, pages = {56--68}, year = {2011},
  doi = {10.1038/nrg2918}
}

@article{guney2016,
  author = {Guney, Emre and Menche, J{\"o}rg and Vidal, Marc and Barab{\'a}si, Albert-L{\'a}szl{\'o}},
  title = {Network-based in silico drug efficacy screening},
  journal = {Nature Communications}, volume = {7}, pages = {10331}, year = {2016},
  doi = {10.1038/ncomms10331}
}

@article{chandak2023,
  author = {Chandak, Payal and Huang, Kexin and Zitnik, Marinka},
  title = {Building a knowledge graph to enable precision medicine},
  journal = {Scientific Data}, volume = {10}, pages = {67}, year = {2023},
  doi = {10.1038/s41597-023-01960-3}
}

@article{liu2011,
  author = {Liu, Yang-Yu and Slotine, Jean-Jacques and Barab{\'a}si, Albert-L{\'a}szl{\'o}},
  title = {Controllability of complex networks},
  journal = {Nature}, volume = {473}, pages = {167--173}, year = {2011},
  doi = {10.1038/nature10011}
}

@article{taubes2021,
  author = {Taubes, Alice and Nova, Phil and Zalocusky, Kelly A. and others},
  title = {Experimental and real-world evidence supporting the computational repurposing of bumetanide for {APOE4}-related {Alzheimer's} disease},
  journal = {Nature Aging}, volume = {1}, pages = {932--947}, year = {2021},
  doi = {10.1038/s43587-021-00122-7}
}

@article{huang2024,
  author = {Huang, Kexin and Chandak, Payal and Wang, Qianwen and others},
  title = {A foundation model for clinician-centered drug repurposing},
  journal = {Nature Medicine}, volume = {30}, pages = {3601--3613}, year = {2024},
  doi = {10.1038/s41591-024-03233-x}
}

@article{li2025deepdrug,
  author = {Li, Victor O. K. and Han, Yang and Kaistha, Tushar and Zhang, Qi and Downey, Jocelyn and Gozes, Illana and Lam, Jacqueline C. K.},
  title = {{DeepDrug} as an expert guided and {AI} driven drug repurposing methodology for selecting the lead combination of drugs for {Alzheimer's} disease},
  journal = {Scientific Reports}, volume = {15}, pages = {2093}, year = {2025},
  doi = {10.1038/s41598-025-85947-7}
}

@article{li2025deepdrug2,
  author = {Li, Victor O. K. and Han, Yang and Lam, Jacqueline C. K. and Downey, Jocelyn},
  title = {{DeepDrug2}: A germline-focused graph neural network framework for {Alzheimer's} drug repurposing validated by electronic health records},
  journal = {medRxiv}, year = {2025},
  doi = {10.1101/2025.05.13.25327570}
}

@article{li2025rgf,
  author = {Li, Victor O. K. and Han, Yang and Lam, Jacqueline C. K.},
  title = {Unravelling causal genetic biomarkers of {Alzheimer's} disease via neuron to gene-token backtracking in neural architecture: A groundbreaking {Reverse-Gene-Finder} approach},
  journal = {Proceedings of the AAAI Conference on Artificial Intelligence},
  volume = {39}, number = {17}, pages = {18466--18475}, year = {2025},
  doi = {10.1609/aaai.v39i17.34032}
}

@article{liberzon2011,
  author = {Liberzon, Arthur and Subramanian, Aravind and Pinchback, Reid and Thorvaldsd{\'o}ttir, Helga and Tamayo, Pablo and Mesirov, Jill P.},
  title = {Molecular signatures database ({MSigDB}) 3.0},
  journal = {Bioinformatics}, volume = {27}, number = {12}, pages = {1739--1740}, year = {2011},
  doi = {10.1093/bioinformatics/btr260}
}

@article{kanehisa2025,
  author = {Kanehisa, Minoru and Furumichi, Miho and Sato, Yoko and Matsuura, Yuriko and Ishiguro-Watanabe, Mari},
  title = {{KEGG}: biological systems database as a model of the real world},
  journal = {Nucleic Acids Research}, volume = {53}, number = {D1}, pages = {D672--D677}, year = {2025},
  doi = {10.1093/nar/gkae909}
}

@misc{msigdb2026,
  author = {{GSEA--MSigDB Team}},
  title = {{MSigDB} {2026.1.Hs} release notes}, year = {2026},
  url = {https://docs.gsea-msigdb.org/MSigDB/Release_Notes/MSigDB_2026.1.Hs/},
  note = {Human collections; accessed 2 October 2026}
}

@article{wang2012,
  author = {Wang, Lei and Chiang, Hsueh-Cheng and Wu, Wenjuan and others},
  title = {Epidermal growth factor receptor is a preferred target for treating amyloid-{$\beta$}-induced memory loss},
  journal = {Proceedings of the National Academy of Sciences USA}, volume = {109}, number = {41}, pages = {16743--16748}, year = {2012},
  doi = {10.1073/pnas.1208011109}
}

@article{iwata2001,
  author = {Iwata, N. and Tsubuki, S. and Takaki, Y. and others},
  title = {Metabolic regulation of brain {A}{$\beta$} by neprilysin},
  journal = {Science}, volume = {292}, number = {5521}, pages = {1550--1552}, year = {2001},
  doi = {10.1126/science.1059946}
}

@article{menche2015,
  author = {Menche, J{\"o}rg and Sharma, Amitabh and Kitsak, Maksim and Ghiassian, Susan Dina and Vidal, Marc and Loscalzo, Joseph and Barab{\'a}si, Albert-L{\'a}szl{\'o}},
  title = {Uncovering disease-disease relationships through the incomplete interactome},
  journal = {Science}, volume = {347}, number = {6224}, pages = {1257601}, year = {2015},
  doi = {10.1126/science.1257601}
}

@article{yu2007,
  author = {Yu, Haiyuan and Kim, Philip M. and Sprecher, Emmett and Trifonov, Valery and Gerstein, Mark},
  title = {The importance of bottlenecks in protein networks: correlation with gene essentiality and expression dynamics},
  journal = {PLoS Computational Biology}, volume = {3}, number = {4}, pages = {e59}, year = {2007},
  doi = {10.1371/journal.pcbi.0030059}
}

@article{cheng2018,
  author = {Cheng, Feixiong and Desai, Rishi J. and Handy, Diane E. and Wang, Ruisheng and Schneeweiss, Sebastian and Barab{\'a}si, Albert-L{\'a}szl{\'o} and Loscalzo, Joseph},
  title = {Network-based approach to prediction and population-based validation of in silico drug repurposing},
  journal = {Nature Communications}, volume = {9}, pages = {2691}, year = {2018},
  doi = {10.1038/s41467-018-05116-5}
}

@article{himmelstein2017,
  author = {Himmelstein, Daniel Scott and Lizee, Antoine and Hessler, Christine and others},
  title = {Systematic integration of biomedical knowledge prioritizes drugs for repurposing},
  journal = {eLife}, volume = {6}, pages = {e26726}, year = {2017},
  doi = {10.7554/eLife.26726}
}

@article{liu2019carnival,
  author = {Liu, Anika and Trairatphisan, Panuwat and Gjerga, Enio and Didangelos, Athanasios and Barratt, Jonathan and Saez-Rodriguez, Julio},
  title = {From expression footprints to causal pathways: contextualizing large signaling networks with {CARNIVAL}},
  journal = {npj Systems Biology and Applications}, volume = {5}, pages = {40}, year = {2019},
  doi = {10.1038/s41540-019-0118-z}
}

@article{cheng2019,
  author = {Cheng, Feixiong and Kov{\'a}cs, Istv{\'a}n A. and Barab{\'a}si, Albert-L{\'a}szl{\'o}},
  title = {Network-based prediction of drug combinations},
  journal = {Nature Communications}, volume = {10}, pages = {1197}, year = {2019},
  doi = {10.1038/s41467-019-09186-x}
}

@article{sims2017,
  author = {Sims, Rebecca and van der Lee, Sven J. and Naj, Adam C. and others},
  title = {Rare coding variants in {PLCG2}, {ABI3}, and {TREM2} implicate microglial-mediated innate immunity in {Alzheimer's} disease},
  journal = {Nature Genetics}, volume = {49}, number = {9}, pages = {1373--1384}, year = {2017},
  doi = {10.1038/ng.3916}
}

@article{solomon2022,
  author = {Solomon, Shiden and Sampathkumar, Nirmal Kumar and Carre, Ivo and others},
  title = {Heterozygous expression of the {Alzheimer's} disease-protective {PLC}{$\gamma$}{2 P522R} variant enhances {A}{$\beta$} clearance while preserving synapses},
  journal = {Cellular and Molecular Life Sciences}, volume = {79}, number = {8}, pages = {453}, year = {2022},
  doi = {10.1007/s00018-022-04473-1}
}

@article{chapuis2013,
  author = {Chapuis, J. and Hansmannel, F. and Gistelinck, M. and others},
  title = {Increased expression of {BIN1} mediates {Alzheimer} genetic risk by modulating tau pathology},
  journal = {Molecular Psychiatry}, volume = {18}, number = {11}, pages = {1225--1234}, year = {2013},
  doi = {10.1038/mp.2013.1}
}

@article{wahrle2005,
  author = {Wahrle, Suzanne E. and Jiang, Hong and Parsadanian, Maia and Hartman, Richard E. and Bales, Kelly R. and Paul, Steven M. and Holtzman, David M.},
  title = {Deletion of {Abca1} increases {A}{$\beta$} deposition in the {PDAPP} transgenic mouse model of {Alzheimer} disease},
  journal = {Journal of Biological Chemistry}, volume = {280}, number = {52}, pages = {43236--43242}, year = {2005},
  doi = {10.1074/jbc.M508780200}
}

@article{mathys2019,
  author = {Mathys, Hansruedi and Davila-Velderrain, Jose and Peng, Zhuyu and others},
  title = {Single-cell transcriptomic analysis of {Alzheimer's} disease},
  journal = {Nature}, volume = {570}, number = {7761}, pages = {332--337}, year = {2019},
  doi = {10.1038/s41586-019-1195-2}
}

@article{shannon1948,
  author = {Shannon, Claude E.},
  title = {A mathematical theory of communication},
  journal = {The Bell System Technical Journal}, volume = {27}, number = {3}, pages = {379--423}, year = {1948},
  doi = {10.1002/j.1538-7305.1948.tb01338.x}
}
